\documentclass[aps,prd,twocolumn,amsmath,showpacs,superscriptaddress,nofootinbib,nopreprintnumbers,tightenlines,notitlepage]{revtex4-1}

\usepackage{verbatim}
\usepackage[T1]{fontenc}
\usepackage[utf8]{inputenc}
\usepackage[american]{babel}
\usepackage{epsfig}
\usepackage{graphicx}

\usepackage{hyperref}

\usepackage{booktabs}
\usepackage{multirow}
\usepackage{dcolumn}
\usepackage{amsmath}
\usepackage{mathtools}
\usepackage{amsfonts}
\usepackage{amssymb}
\usepackage{epstopdf}
\usepackage{bm}
\usepackage{siunitx}
\usepackage{braket}
\usepackage{enumitem}
\usepackage{soul}
\usepackage[table]{xcolor}
\usepackage{color}
\usepackage{transparent}

\usepackage{enumitem}
\usepackage{graphicx}
\usepackage[font=small]{caption}
\usepackage{subcaption}
\usepackage{pifont}

\usepackage{todonotes}

\begin{document}

\title{Model marginalized constraints on neutrino properties from cosmology}

\author{Eleonora Di Valentino}
\email{e.divalentino@sheffield.ac.uk}
\affiliation{School of Mathematics and Statistics, University of Sheffield, Hounsfield Road, Sheffield S3 7RH, United Kingdom}

\author{Stefano Gariazzo}
\email{gariazzo@to.infn.it}
\affiliation{Istituto Nazionale di Fisica Nucleare (INFN), Sezione di Torino, Via P.\ Giuria 1, I-10125 Turin, Italy}

\author{Olga Mena}
\email{omena@ific.uv.es}
\affiliation{Instituto de F{\'\i}sica Corpuscular  (CSIC-Universitat de Val{\`e}ncia), E-46980 Paterna, Spain}

\newcommand{\lcdm}{$\Lambda$CDM}
\newcommand{\Neff}{\ensuremath{N_{\text{eff}}}}
\newcommand{\mnu}{\ensuremath{\sum m_\nu}}
\newcommand{\dmsol}{\ensuremath{\Delta m^2_{21}}}
\newcommand{\dmatm}{\ensuremath{|\Delta m^2_{31}|}}
\newcommand{\sg}[1]{\textcolor{blue}{#1}}

\begin{abstract}
We present robust, model-marginalized limits on both the total neutrino mass ($\mnu$) and abundances (\Neff) to minimize the role of parameterizations, priors and models when extracting neutrino properties from cosmology.
The cosmological observations we consider are Cosmic Microwave Background temperature fluctuation and polarization measurements, Supernovae Ia luminosity distances, Baryon Acoustic Oscillation observations and determinations of the growth rate parameter from the Data Release 16 of the Sloan Digital Sky Survey IV.
The degenerate neutrino mass spectrum (which implies the prior $\mnu>0$) is weakly or moderately preferred over the normal and inverted hierarchy possibilities,  which imply the priors $\mnu>0.06$ and $\mnu>0.1$~eV respectively.
Concerning the underlying cosmological model, the \lcdm\ minimal scenario is almost always strongly preferred over the possible extensions explored here.
The most constraining $95\%$~CL bound on the total neutrino mass in the \lcdm+$\mnu$ picture is $\mnu< 0.087$~eV.
The parameter \Neff\ is restricted to $3.08\pm 0.17$ ($68\%$~CL) in the \lcdm+\Neff\ model.
These limits barely change when considering the \lcdm + $\mnu$ + \Neff\ scenario.
Given the robustness and the strong constraining power of the cosmological measurements employed here, the model-marginalized posteriors obtained considering a large spectra of non-minimal cosmologies are very close to the previous bounds, obtained within the \lcdm\ framework in the degenerate neutrino mass spectrum.
Future cosmological measurements may improve the current Bayesian evidence favouring the degenerate neutrino mass spectra, challenging therefore the consistency between cosmological neutrino mass bounds and oscillation neutrino measurements, and potentially suggesting a more complicated cosmological model and/or neutrino sector.

\end{abstract}

\maketitle

\section{Introduction}
Cosmological measurements are continuously improving the limits on the fraction of the energy's budget in the universe in the form of massive neutrinos in an impressive way: the tightest limit to date within the most economical \lcdm\ scenario is $\mnu <0.09$~eV at $95\%$~CL~\cite{Palanque-Delabrouille:2019iyz,DiValentino:2021hoh} based on the combination of Planck temperature, polarization and lensing observations~\cite{Planck:2018vyg,Planck:2019nip} and SDSS-IV data~\cite{Bautista:2020ahg,Gil-Marin:2020bct}.
Despite the fact that the aforementioned limit is based on the minimal \lcdm\ model, current cosmological limits on $\mnu$ should be regarded as robust, in the sense that they are difficult to avoid in simple extensions of the standard \lcdm, e.g.\ models relaxing the flatness assumption or the inflationary parameters, using the Planck data.
Indeed, by relaxing the cosmological constant -minimal dark energy scenario- the neutrino mass bounds could even get tighter~\cite{Vagnozzi:2018jhn,RoyChoudhury:2018gay}: the $95\%$~CL limit would be $\mnu <0.08$~eV in this case. 
There are however some other (not so simple and straightforward) extensions of the \lcdm\ model which could lead to larger departures from the baseline $95\%$~CL upper bound of $0.09$~eV.
That would be the case of non-standard neutrino physics~\cite{Barenboim:2020vrr,Abellan:2021rfq,Escudero:2020ped,Lorenz:2021alz,Oldengott:2019lke,Alvey:2021sji,Esteban:2021ozz,Mosbech:2020ahp}, exotic dark sectors~\cite{Yang:2020uga,Yang:2020tax,Yang:2020ope,Amendola:2007yx,Stadler:2018dsa}, a mixed hot dark matter scenario~\cite{DEramo:2022nvb,Giare:2020vzo,DiValentino:2015wba,Archidiacono:2015mda,Giusarma:2014zza,Archidiacono:2013cha,Melchiorri:2007cd}, or a modified CMB lensing pattern, parameterized this case by $A_{\rm lens}$~\cite{Renzi:2017cbg,RoyChoudhury:2019hls,Sgier:2021bzf,DiValentino:2021imh,Esteban:2022rjk}, among other possibilities. 

Another source of variability in the neutrino mass bounds is due to the underlying neutrino mass spectrum: the input from neutrino oscillation experiments imposes a lower bound on the neutrino mass which is approximately 0.06~eV (0.10~eV) in the normal (inverted) schemes, characterized by $m_1<m_2<m_3$ and $m_3<m_1<m_2$, respectively.
While current cosmological measurements are unable to extract the precise ordering of the individual neutrino masses~\cite{Gariazzo:2022ahe}, oscillation data only provides weak to moderate preference for the normal mass ordering ($2-2.7\sigma$)~\cite{deSalas:2020pgw,Esteban:2020cvm,Capozzi:2021fjo}.
The upper limit on $\mnu$ will inevitably rely on the lower limit established for the prior, i.e.\ 0, 0.06 and 0.1 eV for the degenerate, normal and inverted neutrino mass spectra respectively.

It is therefore clear that the role of parameterizations, priors and models may lead to different constraints on the cosmological neutrino mass.
Here, following the method of Ref.~\cite{Gariazzo:2018meg}, we exploit Bayesian model comparison techniques to compute model-marginalized cosmological parameter limits on $\mnu$.
The structure of the manuscript is as follows.
We start in Sec.~\ref{sec:analysis} describing the parameterizations, the numerical and the statistical methods and the cosmological observations exploited in our analyses.
Sections \ref{sec:resultsmnu} and \ref{sec:resultsneff} detail the limits obtained within cosmologies in which the total neutrino mass or the effective number of relativistic degrees of freedom are free parameters to be constrained by data.
We perform an exhaustive screening of different parameterizations, priors and models, computing also the marginalized bounds on both $\mnu$ and $\Neff$.
We conclude in Sec.~\ref{sec:conclusions}.

\section{Analysis setup}
\label{sec:analysis}
We shall present in what follows the parameterizations and models explored, detailing our numerical analyses, our statistical method and the cosmological observations used to constrain $\mnu$ and $\Neff$ in a model-marginalized manner.

\subsection{Parameterizations}
\label{subs:param}

As mentioned above, one crucial aspect in parameterizing neutrino masses in cosmological analyses is how one implements the neutrino mass hierarchy.
The three possible standard choices are to consider three degenerate neutrino masses (DH) and simply impose that they are positive ($\mnu>0$), or either implement the constraints that take into account neutrino oscillation results,
thus enforcing the masses to have a normal (NH) or inverted hierarchy (IH)
~\footnote{It is unlikely that present and near-future cosmological measurements will be able to distinguish the details of these scenarios by measuring the effect of the independent mass eigenstates~\cite{Archidiacono:2020dvx}.}.
The difference between these three possibilities is the lower limit on the allowed \mnu\ prior, that neutrino oscillations impose within the NH and IH scenarios:
given the measured values of \dmsol\ and \dmatm, we implement a lower limit of $\mnu>0.06$~eV (NH) or $\mnu>0.1$~eV (IH), respectively.

\subsection{Numerical calculations}
\label{subs:calc}
Cosmological observables are computed with the publicly available Boltzmann solver code \texttt{CAMB}~\cite{Lewis:1999bs}.
To derive bounds on the proposed scenarios, we make use of the efficient and well-known \texttt{CosmoMC}~\cite{Lewis:2002ah}, publicly available at \url{http://cosmologist.info/cosmomc/}, which we use to constrain the cosmological parameters by means of MCMC sampling.
This cosmological package is equipped with a convergence diagnostic following the Gelman and Rubin prescription~\cite{Gelman:1992zz}.
Additionally, the \texttt{CosmoMC} package supports the Planck 2018 likelihood~\cite{Planck:2019nip}.
In order to compute the Bayesian evidences we exploit the \texttt{MCEvidence}~\cite{Heavens:2017afc} tool.

\subsection{Statistical method}
\label{subs:stat}
The goal of this paper is to show the robustness of cosmological results obtained by the most powerful combination in Ref.~\cite{DiValentino:2021hoh}
when extensions of the minimal \lcdm\ model are considered, by using Bayesian evidences to compute a model-marginalized 1D posterior on both \mnu\ and \Neff.
Such model-marginalization technique has been used recently in~\cite{Gariazzo:2018meg},
where Planck 2015 data was exploited to show that a relatively large variation of the \mnu\ limits
was allowed when extending the cosmological model to take into account additional parameters,
such as the universe curvature, \Neff\ or dark energy parameters.

In the following, we shall summarize our statistical method.
Let us consider a set of models, indicated with $\mathcal{M}_i$.
In order to compute the model-marginalized posterior,
one needs the model posterior probabilities $p_i$:
\begin{equation}
\label{eq:modelposterior}
p_i = \frac{\pi_i Z_i}{\sum _j\pi_j Z_j}\,,
\end{equation}
where $\pi_i$ indicates the model prior probability and $Z_i$ its Bayesian evidence (marginal likelihood), and the sum runs over all the considered models~\footnote{Notice that, in principle, infinite models should be considered.
In practice, however, most of the models are disfavored enough by data to be irrelevant in the calculations.}.
The model-marginalized posterior $p(\theta|d)$
for some parameters $\theta$, given some data $d$ is then computed as
\begin{equation}
\label{eq:mmposterior_def}
p(\theta|d)
\equiv
\sum_i
p(\theta|d,\mathcal{M}_i)
p_i
\,,
\end{equation}
where $p(\theta|d,\mathcal{M}_i)$ is the parameter posterior within the model $\mathcal{M}_i$.
Assuming that all the models have the same initial prior $\pi_i$
and using the Bayes factors $B_{i0}=Z_i/Z_0$ with respect to the favored model $\mathcal{M}_0$ within the set,
one can rewrite the previous expression to read\,,
\begin{equation}
\label{eq:mmposterior}
p(\theta|d)
=
\frac{\sum_i
p(\theta|d,\mathcal{M}_i)
B_{i0}
}{
\sum_j B_{j0}
}
\,.
\end{equation}
It is easy to notice that the contribution from each model to the model-marginalized posterior
depends on its Bayes factor with respect to the preferred one,
or in other words on how much the model is efficient in describing the data,
so that if the Bayes factors are large in favor of the simplest model
(which is normally the preferred one),
its extensions will have a small impact on the model-marginalized posterior.

The effect of prior is broader than simply modifying the Bayesian evidences.
In particular, in the case of open likelihoods, such as in the case of the sum of neutrino masses,
which are unconstrained from below,
the posterior results can strongly vary if the prior choice is modified.
One typical example, already introduced in section~\ref{subs:param}, relates to whether one should
consider neutrino oscillation results when imposing the prior range on \mnu:
this choice has an impact on the Bayesian limits one obtains, see e.g.~\cite{Wang:2017htc,RoyChoudhury:2019hls,Vagnozzi:2017ovm}. The prior choice may also have strong impacts on the determination of the mass ordering,
see e.g.~\cite{Schwetz:2017fey,Gariazzo:2018pei,Gariazzo:2022ahe}.

In order to avoid the dependency on prior in determining credible intervals,
one possibility is to compute frequentist-like limits,
based for example on maximum likelihood methods,
which do not require the integration of the posterior over the parameter space.
Another method, introduced more than twenty years ago~\cite{Astone:1999wp,DAgostini:2000edp,DAgostini:2003}
and more recently revived in~\cite{Gariazzo:2019xhx}, considers an analysis method based on Bayes factors
to eliminate the dependence of the bounds on the parameter prior.
Given some model $\mathcal{M}$ which contains a parameter $x$,
the ``relative belief updating ratio'' $\mathcal{R}(x_1, x_2|d,\mathcal{M})$ can be defined as follows:
\begin{equation}
\label{eq:R_def}
\mathcal{R}(x_1, x_2|d,\mathcal{M})
\equiv
\frac{Z^{x_1}_\mathcal{M}}{Z^{x_2}_\mathcal{M}}
\,,
\end{equation}
where the $Z^{x}_\mathcal{M}$
is defined as the Bayesian evidence of model $\mathcal{M}$,
but
fixing $x$ to a specific value%
\footnote{We also assume that the prior on $x$ is independent on the other parameters and viceversa.}:
\begin{equation}
\label{eq:Zx_def}
Z^{x}_\mathcal{M}
=
\int_{\Omega_\psi}
d\psi
\pi(\psi|\mathcal{M})
\mathcal{L}_\mathcal{M}(x,\psi)
\,,
\end{equation}
where $\psi$ represents all the parameters in model $\mathcal{M}$ except $x$,
which can vary in a parameter space $\Omega_\psi$,
$\pi(\psi|\mathcal{M})$ is their prior (notice that the $x$ prior is not included here)
and
$\mathcal{L}$ is the likelihood.

From Eq.~\eqref{eq:R_def}, we easily understand that the relative belief updating ratio
does not represent a probability,
as it is the ratio of two evidences.
This is important because instead of integrating over the parameter space of $x$ in order to obtain a bound,
we simply have to study the function values
in order to determine which are the constraints on $x$.
For example, one can define as strongly disfavored the values of $x$ for which $\mathcal{R}(x_1, x_2|d,\mathcal{M})<T$, where $T$ is a given threshold, e.g.\ $T=\exp(-3)$. From the equation, and taking into account the fact that \eqref{eq:Zx_def} is independent of
the prior on $x$, we also learn that $\mathcal{R}(x_1, x_2|d,\mathcal{M})$ is completely prior-independent.
Moreover, using the Bayes theorem,
it is possible to show that another way to write the function $\mathcal{R}(x_1, x_2|d,\mathcal{M})$ is
\begin{equation}
\label{eq:R}
\mathcal{R}(x_1, x_2|d,\mathcal{M})
=
\frac{
p(x_1|d,\mathcal{M})/\pi(x_1|\mathcal{M})
}{
p(x_2|d,\mathcal{M})/\pi(x_2|\mathcal{M})
}\,,
\end{equation}
where $\pi(x|\mathcal{M})$ is the unidimensional prior on $x$.
This is the most convenient expression to obtain the function $\mathcal{R}$ from a MCMC run.

In the case of open likelihoods, it is particularly convenient to compute
the relative belief updated ratio to compare any value of the parameter $x$
to the reference $x_0$ value towards which the likelihood becomes insensitive to the parameter. In the case of neutrino masses such reference value is zero:  the $\mathcal{R}(\mnu, 0|d)$ function is expected to be one at small values of \mnu,
when data cannot (yet) discriminate between a tiny value of \mnu\ and zero,
and to be suppressed at large \mnu, which data do disfavor at some level of significance. 

Finally, let us notice that the definition of $\mathcal{R}(x_1,x_2|d)$
can be easily extended to perform a model marginalization:
\begin{equation}
\label{eq:mmR_def}
\mathcal{R}(x_1, x_2|d)
\equiv
\frac{
\sum_i Z^{x_1}_{\mathcal{M}_i}\pi(\mathcal{M}_i)
}{
\sum_j Z^{x_2}_{\mathcal{M}_j}\pi(\mathcal{M}_j)
}
\,,
\end{equation}
where now the evidences $Z^{x}_{\mathcal{M}_j}$ are computed within a specific model and
$\pi(\mathcal{M}_j)$ is the model prior.
In order to write $\mathcal{R}(x_1, x_2|d)$ using the parameter prior and posterior,
the simplest assumption is to consider the same prior $\pi(x)$ within all the models.
In such case, Eq.~\eqref{eq:mmR_def} becomes:
\begin{equation}
\label{eq:mmR}
\mathcal{R}(x_1, x_2|d)
=
\frac{
p(x_1|d)/\pi(x_1)
}{
p(x_2|d)/\pi(x_2)
}\,,
\end{equation}
where $p(x|d)$ is the model-marginalized posterior in Eq.~\eqref{eq:mmposterior_def}.

\subsection{Cosmological data}
\label{subs:data}
We describe in what follows the data combination used here, see Ref.~\cite{DiValentino:2021hoh} for more details. The cosmological observations used in the following are:

\begin{itemize}

\item {\bf Planck}: The Cosmic Microwave Background (CMB) temperature and polarization power spectra from the final release of Planck 2018 {\it plikTTTEEE+lowl+lowE}~\cite{Planck:2018vyg,Planck:2019nip}. We also consider the CMB lensing likelihood~\cite{Planck:2018lbu}.  

\item {\bf Pantheon}: Type Ia Supernovae from the Pantheon sample~\cite{Pan-STARRS1:2017jku}, consisting of 1048 data points.

\item {\bf Baryon Acoustic Oscillations (BAO)}~\footnote{For updated limits on $\sum m_\nu$ exploiting full-shape galaxy clustering measurements instead of the information encoded in the geometrical BAO signal, see Ref.~\cite{Tanseri:2022zfe}.}
: We exploit here the Baryon Acoustic  and Redshift Space Distortions measurements from SDSS spectroscopic galaxy and quasar catalogs. More precisely, we
make use of the BOSS~\cite{BOSS:2012dmf} DR12 and the eBOSS~\cite{Dawson:2015wdb} DR16 cosmological observations. The DR12 is exploited in the \emph{BAO only} form, while in the case of DR16 we profit from  the \emph{BAO+RSD} form~\cite{BOSS:2016wmc,eBOSS:2020yzd} for both Luminous Red Galaxies (LRGs),  and Quasars (QSOs). 
\end{itemize}

\section{Results on the sum of neutrino masses}
\label{sec:resultsmnu}

\begin{table*}
\centering
\renewcommand{\arraystretch}{1.5}
\resizebox{\textwidth}{!}{\begin{tabular}{l @{\hspace{0.2 cm}} c @{\hspace{0.4 cm}} c @{\hspace{0.4 cm}} c @{\hspace{0.4 cm}} cccccccccc}
\toprule
\textbf{Cosmological model} & & \textbf{lnBF}   & \boldmath{$\mnu$\textbf{[eV]}} & \boldmath{$\Neff$} & \boldmath{$\Omega_{k}$} & \boldmath{$\alpha_s$} & \boldmath{$r$} & \boldmath{$w_0$}& \boldmath{$w_a$}& \boldmath{$A_{\text{lens}}$}\\
\hline\hline

$+\mnu$
&  DH & 0.0   &$<0.0866$& --& --&--&--&--&--&      -- \\
&  NH & 1.90  &$<0.129$& --& --&--&--&--&  --&      -- \\
&  IH  & 3.37 &$<0.155$& --& --&--&  --&      -- \\
[2ex]

$+\mnu+\Neff$
&  DH & 3.54 &$<0.0968$ & $3.06\pm0.17$ & --&--&--&--&--&       --\\
&  NH & 5.33 &$<0.131$  & $3.11\pm0.17$ & --&--&--& --& --&      --\\
&  IH & 6.85 &$<0.163$  & $3.15\pm0.17$ & --&--&--& --& --&      --\\[2ex]

$+\mnu+\Omega_{k}$
&  DH & 5.05 &$<0.111$& --&$0.0009\pm0.0019$& --&--&--&--&      -- \\
&  NH & 6.48 &$<0.143$& --&$0.0018\pm0.0019$& --&--&--&--&      -- \\
&  IH & 7.72 &$<0.180$& --&$0.0023\pm0.0019$& --&--&--&--&      -- \\[2ex]

$+\mnu+\alpha_s$
&  DH & 5.14 &$<0.0908$&-- &--& $-0.0044\pm0.0066$&--&--&--&       --\\
&  NH & 6.81 &$<0.128$& --&--& $-0.0043\pm0.0067$&--& --& --&      --\\
&  IH & 8.33 &$<0.157$& --&--& $-0.0046\pm0.0067$&--& --& --&      --\\[2ex]

$+\mnu+r$
&  DH & 4.15  &$<0.0898$&--&--& -- &$<0.127$&--&--&            --\\
&  NH & 5.96  &$<0.130$&--&--& -- &$<0.122$&--   &     --&      --\\
&  IH & 7.58  &$<0.156$&--&--& -- &$<0.124$&--   &     --&      --\\[2ex]

$+\mnu+w_0$
&  DH & 3.51   &$<0.139$&--&--& -- &--&$-1.046\pm0.033$&      --&      --\\
&  NH & 4.58 &$<0.165$&--&-- &--&--&$-1.058^{+0.033}_{-0.030}$&      --&      --\\
&  IH & 6.68 &$<0.204$&--&-- &--&--&$-1.070^{+0.038}_{-0.030}$&      --&      --\\[2ex]

$+\mnu+(w_0>-1)$
&  DH & 5.09 &$<0.0848$&--&--& -- &--&$<-0.962$&      --&      --\\
&  NH & 7.16 &$<0.125$&-- &--&--&--&$<-0.967$&      --&      --\\
&  IH & 8.85 &$<0.157$&-- &--&--&--&$<-0.968$&      --&      --\\[2ex]

$+\mnu+w_0+w_a$
&  DH & 4.70 &$<0.224$&--&--& -- &--&$-0.933^{+0.077}_{-0.089}$&      $-0.52^{+0.43}_{-0.29}$&      --\\
&  NH & 5.11 &$<0.248$&--&--& -- &--&$-0.917^{+0.081}_{-0.090}$&      $-0.65^{+0.42}_{-0.30}$&      --\\
&  IH & 6.58 &$<0.265$&--&--& -- &--&$-0.921^{+0.078}_{-0.095}$&      $-0.68^{+0.45}_{-0.31}$&      --\\[2ex]

$+\mnu+A_{\text{lens}}$
&  DH & 3.52 &$<0.166$&--&--& -- &--&--&--&           $1.071^{+0.037}_{-0.043}$\\
&  NH & 4.12 &$<0.189$&--&--& --&--&--&      --&      $1.086^{+0.037}_{-0.043}$\\
&  IH & 4.91 &$<0.216$&--&--& --&--&--&      --&      $1.101^{+0.035}_{-0.041}$\\[2ex]

\hline
model marginalized & DH & -- & $<0.102$ \\

\bottomrule
    \end{tabular}}
    \caption{Bayes factors with respect to the simplest model in the table, constraints at 68\% and upper limits at 95\% CL,
    for the \lcdm+$\mnu$ model and its extensions (see also Tab.~\ref{tabmnuNeff}).}
\label{tabmnu}
\end{table*}

\begin{table*}
\centering
\renewcommand{\arraystretch}{1.5}
\resizebox{\textwidth}{!}{\begin{tabular}{l @{\hspace{0.2 cm}} c @{\hspace{0.4 cm}} c @{\hspace{0.4 cm}} c @{\hspace{0.4 cm}} cccccccccc}
\toprule
\textbf{Cosmological model} &  & \textbf{lnBF} & \boldmath{$\Neff$}  & \boldmath{$\mnu$\textbf{[eV]}} & \boldmath{$\Omega_{k}$} & \boldmath{$\alpha_s$} & \boldmath{$r$} & \boldmath{$m^{\rm eff}_{\nu, s}$\textbf{[eV]}} & \boldmath{$Y_{\rm He}$}& \boldmath{$w_0$}\\
\hline\hline

$+\mnu+\Neff$
&  DH & 0.0  & $3.06\pm0.17$  &$<0.0968$& --&--&--&--&--&       --\\
&  NH & 1.79 & $3.11\pm0.17$  &$<0.131$& --&--&--& --& --&      --\\
&  IH & 3.31 & $3.15\pm0.17$  &$<0.163$  & --&--&--& --& --&      --\\[2ex]

$+\mnu+\Neff+\Omega_{k}$
&  DH & 4.63 & $3.02\pm0.19$   &$<0.112$& $0.0011\pm0.0021$&--&--&--&--&      --\\
&  NH & 6.34 & $3.05\pm0.18$   &$<0.149$& $0.0019\pm0.0021$&--&--&--&--&      --\\
&  IH & 7.34 & $3.06\pm0.19$   &$<0.179$& $0.0025\pm0.0022$&--&--&--&--&      --\\[2ex]

$+\mnu+\Neff+\alpha_s$
&  DH & 4.76 & $2.98\pm0.19$   &$<0.0892$& --&$-0.0057\pm0.0077$&--&--&--&      --\\
&  NH & 6.59 & $3.06\pm0.19$   &$<0.129$& --&$-0.0041\pm0.0075$&--&--&--&      --\\
&  IH & 8.15 & $3.11\pm0.19$   &$<0.161$& --&$-0.0036\pm0.0076$&--&--&--&      --\\[2ex]

$+\mnu+\Neff+m^{\rm eff}_{\nu, s}$
&  DH & 4.16 & $<3.39$  &$<0.0901$& --  & -- & -- &$<0.304$&--&      --\\
&  NH & 6.06 & $<3.43$ & $<0.130$ & -- & -- &--&$<0.280$&--&--\\
&  IH & 7.72 & $<3.45$ & $<0.159$ & -- & -- &--&$<0.247$&--&--\\[2ex]


$+\mnu+\Neff+w_0$
&  DH & 3.11 & $2.99\pm0.18$   &$<0.124$& --&--&--&--&--&$-1.047\pm0.034$\\
&  NH & 4.39 & $3.00\pm0.18$   &$<0.165$& --&--&--&--&--&$-1.061\pm0.034$\\
&  IH & 5.24 & $3.02\pm0.18$   &$<0.194$& --&--&--&--&--&$-1.071\pm0.035$\\[2ex]

$+\mnu+\Neff+(w_0>-1)$
&  DH & 5.23 & $3.08\pm0.17$   &$<0.0894$& --&--&--&--&--&$<-0.959$\\
&  NH & 6.90 & $3.14\pm0.17$   &$<0.130$& --&--&--&--&--&$<-0.963$\\
&  IH & 8.58 & $3.17\pm0.17$   &$<0.159$& --&--&--&--&--&$<-0.967$\\[2ex]

\hline
model marginalized & DH & -- & $3.05^{+0.19}_{-0.18}$ & $<0.100$\\

\bottomrule
    \end{tabular}}
    \caption{Bayes factors with respect to the simplest model in the table, constraints at 68\% and upper limits at 95\% CL,
    for the \lcdm+$\mnu$+$\Neff$ model and its extensions.}
\label{tabmnuNeff}
\end{table*}

Results obtained in simple extensions of \lcdm+\mnu\ are shown in Tab.~\ref{tabmnu},
for the DH, the NH and the IH parameterizations~\footnote{It is very important to assess the reliability of our results. The bounds found here within the DH case agree with those presented in Ref.~\cite{DiValentino:2021hoh}, which exploits the very same data combination.}. 
From what regards parameterizations of the neutrino mass spectrum pattern, the upper bounds on \mnu\ depend on the assumed hierarchy, with higher bounds corresponding to the IH case, as the parameter space is shifted towards higher values of \mnu\ by the prior limits.
The tightest bounds are instead obtained within the DH case. 
Notice, from the second column of Tab.~\ref{tabmnu}, which shows the values of the Bayes factors, that the DH spectrum is always favoured with respect to the oscillation-driven schemes:
the preference for the DH spectrum is weak or moderate
when confronted against the NH and IH possibilities, respectively.
Concerning the underlying cosmological model, the \lcdm\ minimal scenario is almost always strongly preferred over the possible extensions explored here. 

Interestingly, the most constraining bound is not obtained within the \lcdm+\mnu\ DH case (which is 0.087~eV at 95\% CL), but within the  \lcdm +\mnu+ $w_0>-1$ DH scenario ($\mnu < 0.085$~eV at 95\% CL).
It is well known that when the cosmological constant assumption is relaxed and the possibility of a general dark energy fluid with an equation of state is adopted instead, the constraints on the neutrino mass are strengthened if restricting the parameter space to the \emph{physical} region, i.e.\ to $w_0>-1$~\cite{Vagnozzi:2018jhn,RoyChoudhury:2018gay}.
The less constraining case for the DH parameterization is the \lcdm+\mnu+$w_0$+$w_a$ model, for which the 95\% CL limit exceeds 0.22~eV (DH).
Concerning the mean values and bounds on the other parameters, as \Neff, notice that these two parameters are not strongly correlated, as the error on \Neff\ ($0.17$ at $68\%$~Cl) are identical to those obtained when extracting only \Neff\ (see the first row of Tab.~\ref{tabNeff}).
The limits on the inflationary parameters $\alpha_s$ and $r$ are almost identical to those obtained within the minimal \lcdm\ picture (see also Ref.~\cite{Gerbino:2016sgw}), and the same is true for the curvature of the universe $\Omega_k$ (see also Ref.~\cite{RoyChoudhury:2019hls}) when BAO and Planck are combined together~\footnote{See Refs.~\cite{DiValentino:2019qzk,Handley:2019tkm,Vagnozzi:2020rcz} for a discussion about the BAO and Planck compatibility when the curvature is allowed to vary.}.
On the other hand, it is well-known that there exists a degeneracy between the dark energy equation of state $w$ and the sum of the three active neutrino masses $\mnu$~\cite{Vagnozzi:2017ovm,Archidiacono:2013lva,Giusarma:2013pmn,Hannestad:2004nb,Lesgourgues:2006nd,Hannestad:2010kz,Wong:2011ip,Lesgourgues:2012uu,Archidiacono:2017tlz,Lattanzi:2017ubx,Hannestad:2005gj,Goobar:2006xz,Joudaki:2012fx,Lorenz:2017fgo,Lorenz:2017iez}.
Cosmological neutrino mass bounds become weaker if the dark
energy equation of state is taken as a free parameter which can take values also in the phantom region, i.e.\ $w<-1$.
The reason for this is due to the fact that if $w$ is allowed to vary, the dark matter energy density 
can take very high values and therefore a larger neutrino mass would be allowed.

Finally, when considering the purely phenomenological parameter $A_{\rm lens}$, a positive degeneracy is expected between the former parameter and $\mnu$: an increase in the neutrino mass will imply a depletion in the CMB lensing signal, as neutrinos are hot relics which suppress structure formation at small scales.
This reduction can be compensated by increasing the $A_{\rm lens}$ parameter, which controls the amount of smoothing induced in the CMB peaks by lensing~\cite{Renzi:2017cbg,Capozzi:2017ipn,RoyChoudhury:2019hls,Sgier:2021bzf,DiValentino:2021imh,Esteban:2022rjk}.
The last row of Tab.~\ref{tabmnu} depicts the model marginalized~\footnote{Notice that this limit is computed by considering also the models reported in table~\ref{tabmnuNeff}.}
value of $\mnu$ for the DH case for the different models explored here.
Interestingly, this marginalized limit is extremely tight, $\mnu <0.102$~eV at $95\%$~CL, clearly stating the robustness of cosmological neutrino mass limits and the minor role of the assumed underlying cosmology when considering very powerful and constraining cosmological data sets, as those exploited here. 

\begin{figure*}
\centering
\includegraphics[width=0.7\textwidth]{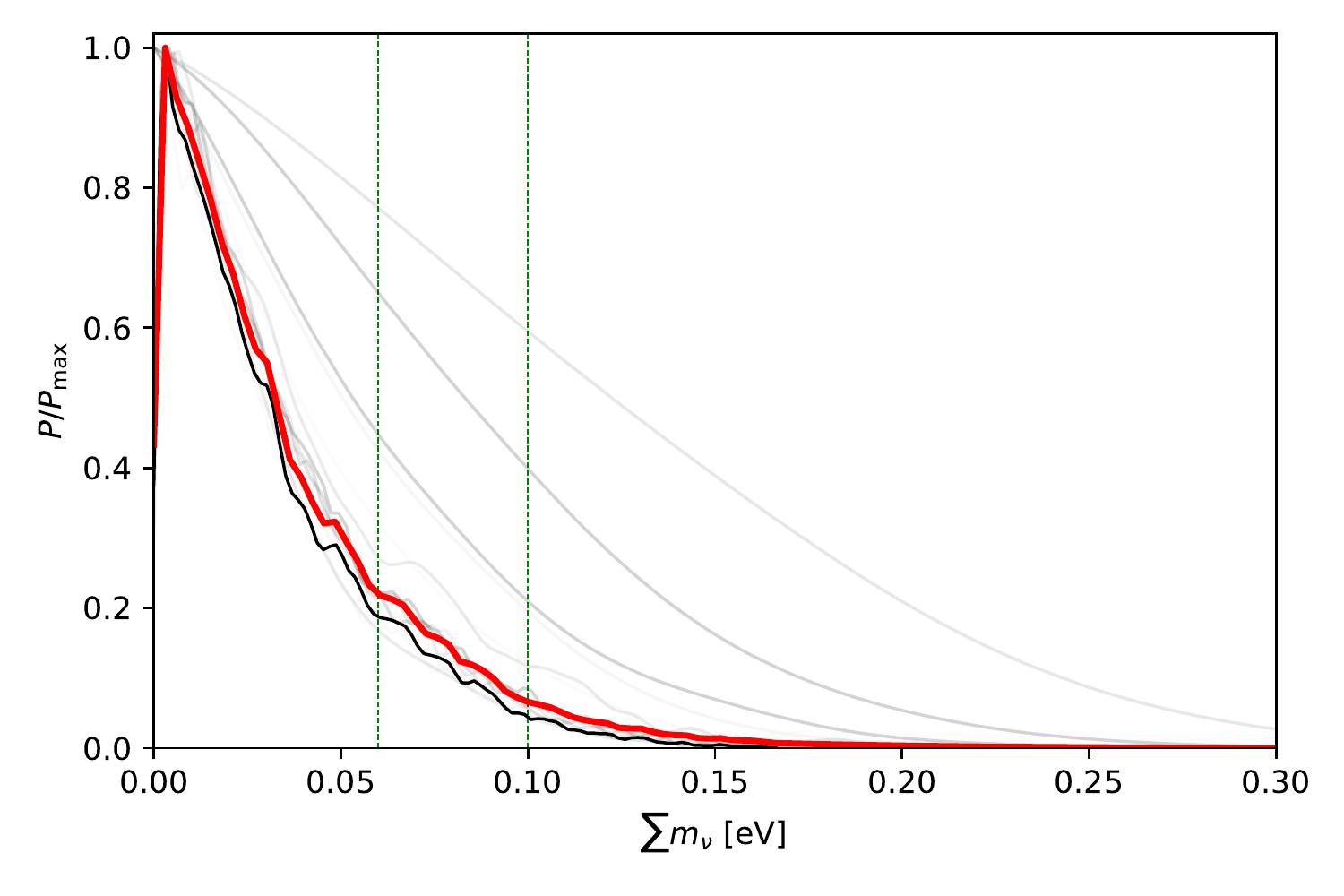}
\caption{
\label{fig:mm_mnu_mnu}
Model-marginalized 1D posterior for $\mnu$ (in red), considering the extensions of the \lcdm+$\mnu$ model from Tabs.~\ref{tabmnu} and \ref{tabmnuNeff}.
Black and gray lines show the 1D posterior within each model, where the darker lines are those that contribute most to the model marginalized posterior, i.e.\ they have the best Bayesian evidences.
Vertical lines indicate the values 0.06 and 0.1 eV, corresponding to the approximate lower limits for \mnu\ within the NH and IH cases.
}
\end{figure*}

The situation is qualitatively similar if we consider the \lcdm+\mnu+\Neff\ model and its extensions,
as reported in Tab.~\ref{tabmnuNeff}, although of course the constraints are slightly relaxed
by the additional freedom that the \Neff\  parameter brings (in particular, when the primordial Helium-4 fraction is considered to be a free parameter, see the next section). In this extended model, the most constraining bound on the total neutrino mass is not obtained within the minimal \lcdm\ framework (in which it reads $\mnu< 0.097$~eV at $95\%$~CL) but in the \lcdm +\mnu+\Neff+ $\alpha_s$ case, in which the $95\%$~CL upper bound is $\mnu< 0.089$~eV within the DH neutrino mass spectrum.
The loosest limit for each of the three neutrino mass spectra is instead always found within the \lcdm +\mnu+\Neff+$w_0$ model, in which $\mnu< 0.124$, $0.165$ and $0.194$~eV at $95\%$~CL in the DH, NH and IH schemes respectively.
As in the \lcdm+\mnu\ cosmology, we also report in the last row of Tab.~\ref{tabmnuNeff} the model marginalized values of both $\mnu$ ( $\mnu <0.1$~eV at $95\%$~CL) and  \Neff\ ($3.05^{+0.19}_{-0.18}$, errors at $68\%$~CL) for the DH case.
These values provide again a strong assessment of the robustness of the cosmological data sets used here and ensure the reliability of the cosmological limits on neutrino properties, as they are extremely stable and tight.

\begin{figure*}
\centering
\includegraphics[width=0.7\textwidth]{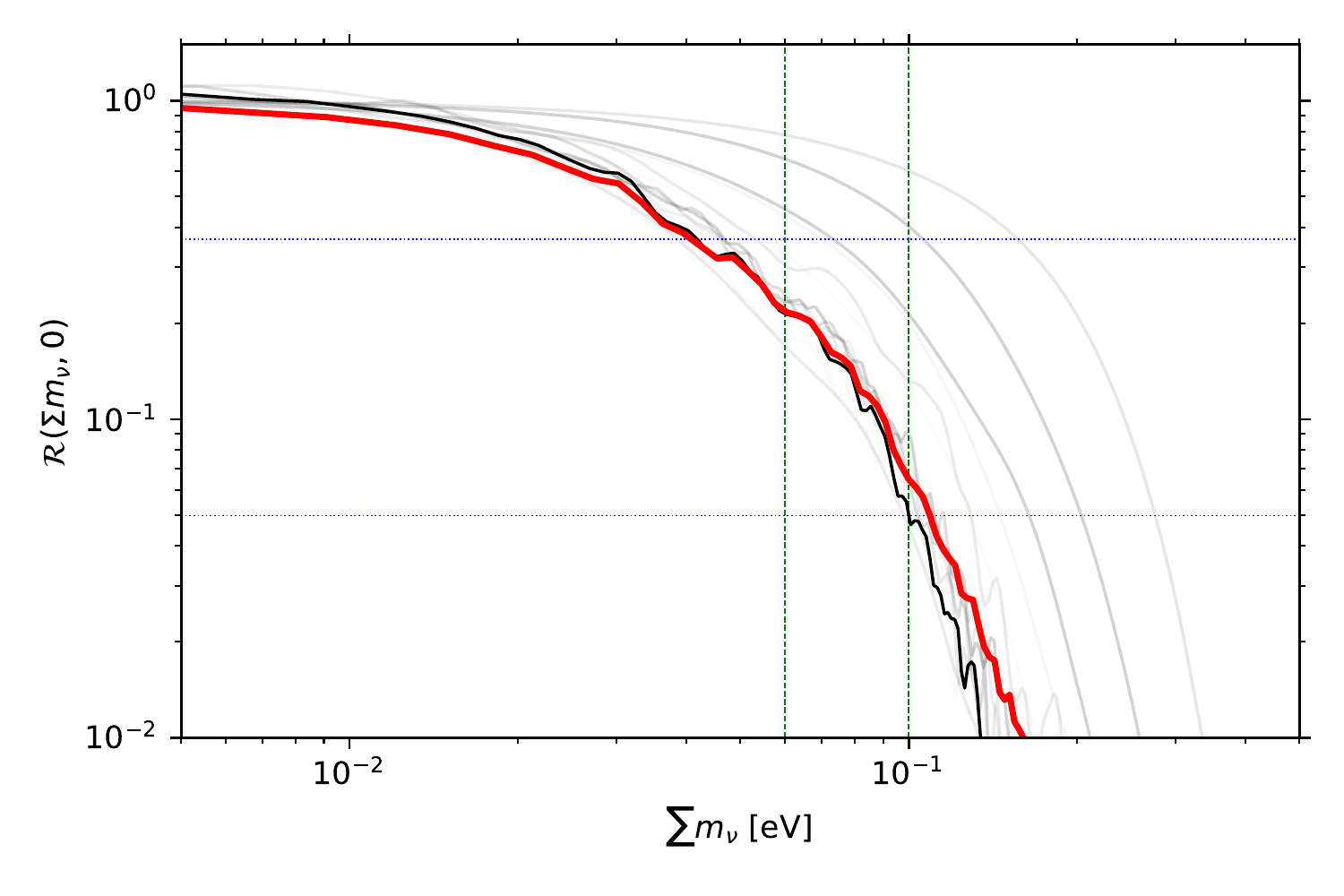}
\caption{
\label{fig:Rmm_mnu_mnu}
Model-marginalized relative belief updating ratio $\mathcal{R}$~\cite{Astone:1999wp,DAgostini:2000edp,DAgostini:2003,Gariazzo:2019xhx} for $\mnu$,
considering the extensions of the \lcdm+$\mnu$ model from Tabs.~\ref{tabNeff} and \ref{tabmnuNeff}.
Black and gray lines show the $\mathcal{R}$ function within each model, where the darker lines are those that contribute most to the model marginalization, i.e.\ they have the best Bayesian evidences.
Vertical lines indicate the values 0.06 and 0.1 eV, corresponding to the approximate lower limits for \mnu\ within the NH and IH cases.
Horizontal lines show the significance levels $\exp(-1)$ and $\exp(-3)$.
}
\end{figure*}

Figure~\ref{fig:mm_mnu_mnu} illustrates the 1D posteriors obtained within each model in Tab.~\ref{tabmnu} and~\ref{tabmnuNeff}, in various scales of gray depicting the Bayes factors
(the black line is the simplest and most favored model, \lcdm+\mnu),
together with the model-marginalized posterior obtained using Eq.~\eqref{eq:mmposterior} (in red).
With respect to previous results~\cite{Gariazzo:2018meg}, we can notice from the tables that
the Bayesian evidences now strongly disfavor extended models with respect to the simplest one, as the constraining power of the data combination exploited here clearly supersedes that of the one employed in Ref.~\cite{Gariazzo:2018meg}, which restricted exclusively to the CMB measurements from the Planck satellite 2015 data release. 
The crucial message is that the model-marginalized posterior is much closer to the one obtained
within the simplest 7-parameters extension of the \lcdm\ and that taking into account additional extensions of the simplest cosmological model has therefore very little impact on the bounds on $\sum m_\nu$: current cosmological neutrino bounds are robust and little model dependent.

Figure \ref{fig:Rmm_mnu_mnu} shows the $\mathcal{R}$ function, Eqs.~(\ref{eq:R}) and (\ref{eq:mmR}), within each model.
This quantity is independent
of the shape and normalization of the prior and it is statistically equivalent to a Bayes factor between a model where  $\mnu$ has been fixed to some value and one where $\mnu=0$. Since cosmological observations are insensitive to the value of $\mnu$ when this quantity is very
small, the function $\mathcal{R}$ is expected to be equal to one for small values of $\mnu$, while it decreases when
large values of $\mnu$ become disfavored. Similarly to the  Bayes factor case, one can compare the limits within each possible extension of the \lcdm\ scenario
by means of the Jeffreys’ scale \cite{Jeffreys:1961a,Trotta:2008qt}.
The horizontal lines in Fig.~\ref{fig:Rmm_mnu_mnu} show the values $\ln R = -1$ and $-3$,
which separate regions where the significance is none,
weak and moderate, accordingly to the Jeffreys’ scale we adopt here, from \cite{Trotta:2008qt}.
Notice that for the case of models which have the best Bayesian evidences (see darker lines), the function $\mathcal{R}$ decreases much faster,  disfavouring moderately values of $\mnu > 0.1$~eV.
In extended models, represented by the lighter curves and contributing less to the model marginalization, the value $\mnu = 0.1$~eV is either unable to be disfavoured or only weakly disfavoured.

\section{Results on the effective number of relativistic species}
\label{sec:resultsneff}

\begin{table*}
\centering
\renewcommand{\arraystretch}{1.5}
\resizebox{\textwidth}{!}{\begin{tabular}{l @{\hspace{0.2 cm}} c @{\hspace{0.4 cm}} c @{\hspace{0.4 cm}} c @{\hspace{0.4 cm}} cccccccccc}
\toprule
\textbf{Cosmological model} &   & \textbf{lnBF} & \boldmath{$\Neff$} & \boldmath{$\mnu$\textbf{[eV]}}  & \boldmath{$\Omega_{k}$} & \boldmath{$\alpha_s$}  & \boldmath{$m^{\rm eff}_{\nu, s}$\textbf{[eV]}} & \boldmath{$Y_{\rm He}$}& \boldmath{$w_0$}& \boldmath{$w_a$}\\
\hline\hline

$+\Neff$ & & 0.0    & $3.08\pm0.17$  &--  &--&--&--&--&--&--\\[2ex]

$+\Neff+\mnu$
&  DH & 3.54 & $3.06\pm0.17$  &$<0.0968$& --&--&--&--&--&       --\\
&  NH & 5.34 & $3.11\pm0.17$  &$<0.131$& --&--&--& --& --&      --\\
&  IH & 6.86  & $3.15\pm0.17$&$<0.163$  & --&--&--& --& --&      --\\[2ex]

$+\Neff+\Omega_{k}$ &  & 4.83  & $3.04\pm0.19$  &-- & $0.0015\pm0.0020$&-- &--&--&--&      -- \\[2ex]

$+\Neff+\alpha_s$ & & 5.00    & $3.03\pm0.19$ &-- &--&$-0.0051\pm0.0074$&--&--& --&      -- \\[2ex]

$+\Neff+m^{\rm eff}_{\nu, s}$ & & 4.39    & $<3.41$ & -- &-- &--&$<0.249$&--&--&     --\\[2ex]

$+\Neff+Y_{\rm He}$ & & 2.26    & $3.17^{+0.27}_{-0.31}$ & -- &-- &--&--&$0.240\pm0.018$&--&--\\[2ex]

$+\Neff+w_0$ & & 2.95    & $2.99\pm0.18$  & -- &-- &--&--&--&$-1.052\pm0.033$&--\\[2ex]

$+\Neff+(w_0>-1)$ & & 5.21    & $3.12\pm0.17$  &-- &-- &--&--&--&$<-0.963$& --\\[2ex]

$+\Neff+w_0+w_a$ & & 3.80    & $2.91\pm0.18$ & -- &-- &--&--&--&$-0.934\pm0.083$&$-0.51^{+0.36}_{-0.30}$\\[2ex]

\hline
model marginalized & & -- & $3.07^{+0.19}_{-0.18}$ \\

\bottomrule
    \end{tabular}}
    \caption{Bayes factors with respect to the simplest model in the table, constraints at 68\% and upper limits at 95\% CL,
    for the \lcdm+$\Neff$ model and its extensions (see also Tab.~\ref{tabmnuNeff}).}
\label{tabNeff}
\end{table*}

The results on $\Neff$, shown in Tab.~\ref{tabNeff}, are qualitatively different from those on $\mnu$, since the likelihood is closed and it is possible to constrain this parameter to be different from zero. 
We only illustrate the DH, NH and IH cases when considering the \Neff\ + $\mnu$ extension.
Notice that the preferred values are shifted towards higher \Neff\ in the IH case with respect to the DH or NH ones, the reason for that there is a (very mild) degeneracy between \Neff\ and $\mnu$.

From Tab.~\ref{tabNeff} one can notice that the \Neff\ mean values and errors do not vary significantly in the \lcdm\ extensions, 
except when the primordial Helium-4 fraction $Y_{He}$ is free (broader errors) or a sterile neutrino is considered (in this very last case, $\Neff>3.044$ by prior).
It is well-known that constraints on \Neff\ from CMB observations mostly come from their effect on the photon diffusion scale at high multipoles, see Ref.~\cite{Hou:2011ec}, and from the early Integrated Sachs Wolfe effect, see Ref.~\cite{Archidiacono:2013fha}.
However,the primordial Helium-4 fraction also modifies the ratio of damping to sound-horizon scales since it affects the electron number density.
Consequently, there exists a significant degeneracy between \Neff\ and $Y_{He}$, making the error bars approximately a factor of two larger than when the Helium-4 fraction is not allowed to freely vary.
In the case of massive sterile neutrinos with a mass $m^{\rm eff}_{\nu, s}$, there is a new degeneracy and only an upper bound is found for \Neff\, in perfect agreement with Planck 2018 results, despite the fact that our constraints on $m^{\rm eff}_{\nu, s}$ are tighter due to the most constraining power of our data combination.
We also report in Tab.~\ref{tabNeff} the model marginalized value of \Neff\,  (\Neff=$3.07^{+0.19}_{-0.18}$, with $68\%$~CL errors), extremely close to the mean value and errors found in the case in which the neutrino mass is also included as a free parameter in the analysis (see Tab.~\ref{tabmnuNeff}) and also very close to the result within the minimal \lcdm\ cosmology when only \Neff\ is considered to be a free parameter, see the very first row of Tab.~\ref{tabNeff}.

\begin{figure*}
\centering
\includegraphics[width=0.7\textwidth]{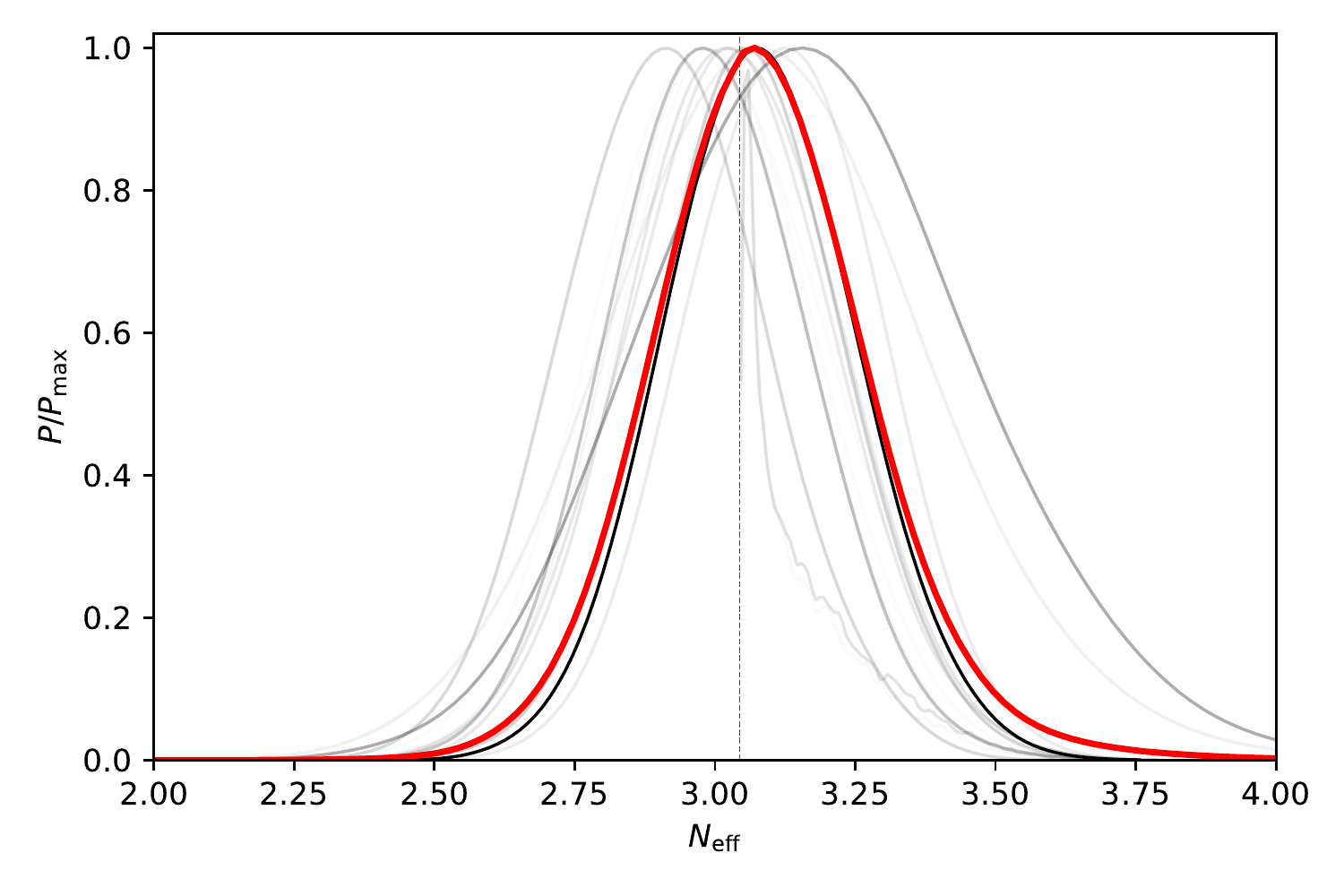}
\caption{
\label{fig:mm_neff_nnu}
Model-marginalized 1D posterior for $\Neff$ (in red), considering the extensions of the \lcdm+$\Neff$ model from Tabs.~\ref{tabNeff} and \ref{tabmnuNeff}.
Black and gray lines show the 1D posterior within each model, where the darker lines are those that contribute most to the model marginalized posterior, i.e.\ they have the best Bayesian evidences.
The vertical line indicates the standard value $\Neff=3.044$~\cite{Akita:2020szl,Froustey:2020mcq,Bennett:2020zkv}.
}
\end{figure*}

In analogy to the previous section, we report the model-marginalized posterior
in Fig.~\ref{fig:mm_neff_nnu},
together with the 1D posteriors within each model, with grayscale colors where black corresponds to the preferred model.
Also in this case, we can see that the model-marginalized posterior (red)
is very close to the one obtained within the \lcdm+\Neff\ model (in black).
Since the likelihood is closed,
we do not show the results in terms of the $\mathcal{R}$ function for \Neff,
as they would be qualitatively equivalent to the standard Bayesian constraints.

\section{Conclusions}
\label{sec:conclusions}

Present bounds on the neutrino masses and abundances, despite their robustness, unavoidably depend on the model, parameterization and prior assumptions.
Namely, constraints tend to relax when the minimal \lcdm\ scenario is extended, and also when the lowest possible value for $\mnu$ is increased from the physical one (i.e.\ $\mnu >0$) to those imposed by neutrino oscillation inputs (i.e.\ $\mnu >0.06$ or $\mnu >0.1$~eV, depending on the precise neutrino mass hierarchy).
In this regard, obtaining model-marginalized limits on both the total neutrino mass ($\mnu$) and abundances (\Neff) are absolutely mandatory to quantify and account for the role of parameterizations, priors and models when extracting neutrino properties from cosmology.
By means of Planck Cosmic Microwave Background temperature fluctuation and polarization measurements, Supernovae Ia luminosity distances, Baryon Acoustic Oscillation observations and determinations of the growth rate parameter from the Data Release 16 of the Sloan Digital Sky Survey IV, we have presented such limits in this manuscript.
Applying Bayesian evidence methods, we find that the  degenerate neutrino mass spectrum is weakly or moderately preferred over the normal and inverted hierarchy possibilities, respectively.
Concerning the underlying cosmological model, the \lcdm\ minimal scenario is almost always strongly preferred over some possible extensions which include a non-vanishing spatial curvature, a time varying dark energy equation of state, a running of the scalar spectral index or a non-standard normalization of the lensing power spectrum, among others.
The most constraining $95\%$~CL bound on the total neutrino mass in the \lcdm+\mnu\ picture is $\mnu< 0.087$~eV in the degenerate neutrino mass spectrum.
The parameter \Neff\ is restricted to $3.08\pm 0.17$ ($68\%$~CL) in the minimal \lcdm+\Neff\ scenario.
These limits barely change when considering simultaneously the \lcdm + $\mnu$ + \Neff\ scenario, or when computing the model-marginalized limits ($\mnu<0.102$~eV and $\Neff=3.07^{+0.19}_{-0.18}$ respectively).
If upcoming cosmological observations keep disfavouring the normal or inverted neutrino mass spectra versus the degenerate case, searches for new physics within the neutrino or the gravitational sectors of the theory may be motivated.

To conclude, the model-marginalized limits on \mnu\ and \Neff\ in the different scenarios explored here are extremely stable and very close to their values in the minimal \lcdm\ scenario.
In particular, our results clearly state the reliability of the observational data sets used here.
In more general grounds, the bounds reported here ensure the robustness of the cosmological limits on neutrino masses and abundances, despite the underlying model assumed for the evolution of our universe.

\begin{acknowledgments}
EDV is supported by a Royal Society Dorothy Hodgkin Research Fellowship.
SG acknowledges financial support from the European Union's Horizon 2020 research and innovation programme under the Marie Skłodowska-Curie grant agreement No 754496 (project FELLINI).
The work of OM is supported by the Spanish grants PID2020-113644GB-I00, PROMETEO/2019/083. 
OM also acknowledges support from the European ITN project HIDDeN (H2020-MSCA-ITN-2019/860881-HIDDeN). 

\end{acknowledgments}

\bibliography{bibliography}

\end{document}